\documentclass[prl,twocolumn,english,superscriptaddress,floatfix,longbibliography]{revtex4}

\usepackage{babel}
\usepackage{graphicx}
\usepackage{bm, amsmath, amssymb}
\usepackage{xcolor}
\usepackage{color}
\usepackage{soul}
\usepackage{pdfpages}
 \usepackage{babel}

\usepackage{amstext}
\usepackage{bbold}
\usepackage{esint}

\newcommand{\new}[1]{\textcolor{black}{#1}}

\newcommand{\mn}[1]{\textcolor{black}{#1}}

\newcommand{\pt}[0]{$\mathcal{PT}$}
\newcommand{\ket}[1]{|{#1}\rangle}
\usepackage{babel}

\begin{document}

\title{Quantum state tomography across the exceptional point in a single dissipative qubit}

\author{M. Naghiloo}
\affiliation{Department of Physics, Washington University, St.~Louis, Missouri 63130}
\author{M. Abbasi}
\affiliation{Department of Physics, Washington University, St.~Louis, Missouri 63130}
\author{Yogesh N. Joglekar}
\affiliation{Department of Physics, Indiana University Purdue University
Indianapolis (IUPUI), Indianapolis, Indiana 46202}
\author{K. W. Murch}
\affiliation{Department of Physics, Washington University, St.~Louis, Missouri 63130}
\affiliation{Institute for Materials Science and Engineering, St.~Louis, Missouri 63130}
\date{\today}

\begin{abstract}

\new{Open systems with gain and loss, described by non-trace-preserving, non-Hermitian Hamiltonians, have been a subject of intense research recently. The effect of exceptional-point degeneracies on the dynamics of classical systems has been observed through remarkable phenomena such as the parity-time symmetry breaking transition, asymmetric mode switching, and optimal energy transfer}. On the other hand, consequences of an exceptional point for quantum evolution and decoherence are hitherto unexplored. Here, we use post-selection on a three-level superconducting transmon circuit with tunable Rabi drive, dissipation, and detuning to carry out quantum state tomography of a single dissipative qubit in the vicinity of its exceptional point. Quantum state tomography reveals the \pt symmetry breaking transition at zero detuning, decoherence enhancement at finite detuning, and a quantum signature of the exceptional point in the qubit relaxation state.  Our observations demonstrate rich phenomena associated with non-Hermitian physics \new{such as non-orthogonality of eigenstates in a} fully quantum regime and open routes to explore and harness exceptional point degeneracies for enhanced sensing and quantum information processing.
\end{abstract}
\maketitle


In introductory treatments of quantum mechanics one typically assumes that a Hamiltonian describing a physical system is Hermitian thus ensuring the reality of energy eigenvalues and a unitary time evolution.  Many open physical systems are instead described by effective non-Hermitian Hamiltonians that characterize the gain or loss of energy or particle-number from the system.  In recent years there has been growing interest in non-Hermitian systems, particularly those with space-time reflection ($\mathcal{PT}$) symmetry exhibiting transitions from purely real to complex-conjugate spectra \cite{bend98,most10}.  Such non-Hermitian systems have been realized with optical \cite{rute10,rege12,hoda14,feng14,peng14_2,feng17,el2018} and mechanical systems \cite{bendpend} with balanced gain and loss, or with mode-selective loss \cite{guo09,zeun15,li16,weim16,xiao17}. The degeneracies of such Hamiltonians occur at exceptional points (EPs) where the eigenvalues, and corresponding eigenmodes coalesce, and are topological in nature~\cite{dopp16,zhen15, gao15,kato,heis12,zhan17}. Open systems in the vicinity of EPs have shown functionalities including lasing \cite{peng14,miao16,wong16}, topological features \cite{wang09,rech13, chan14,el2018}, optimal energy transfer \cite{xu16, assa17}, and enhanced sensing \cite{hoda17,chen17} that are absent in their closed counterparts. However, most of these realizations are limited to classical (wave) systems in which the amplitude information is measured, but the phase information is ignored. Thus, the effects of the \pt-transition and a system's proximity to the EP on its full quantum evolution and decoherence are open questions.


Here, we employ bath engineering techniques to realize a superconducting circuit with quantum energy levels that are  described by a non-Hermitian Hamiltonian and use quantum state tomography to observe the dynamics of the non-Hermitian qubit across and in the vicinity of the exceptional point. At zero detuning, we observe the $\mathcal{PT}$ symmetry breaking transition as manifested in the evolution of both diagonal and off-diagonal elements of the system's density matrix. \new{ By measuring the overlap of the eigenstates of the effective Hamiltonian across the $\mathcal{PT}$ transition, we observe their coalescence at the EP.} We go on to show that the decoherence rate of the qubit and the steady state it reaches are both affected by the system's proximity to the EP.

\begin{figure*}
\centering
\includegraphics[width = 0.88\textwidth]{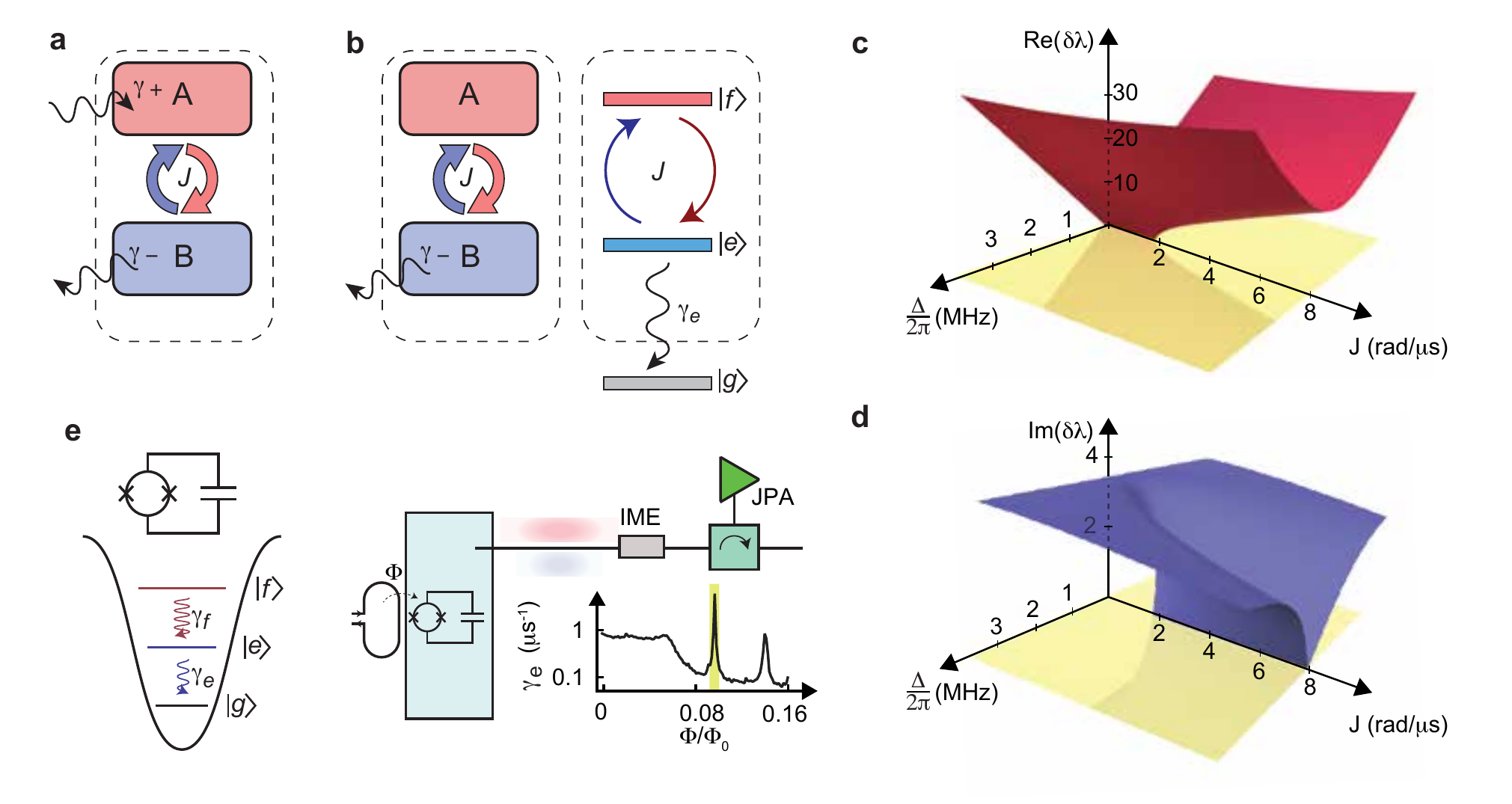}
\caption{  \new{{\bf Experimental overview}}.  (a) A system with balanced gain and loss exhibits \pt symmetry. (b) Systems with mode selective losses, where one mode exhibits loss, manifest the same topological features as \pt-symmetric systems with balanced gain and loss.  We realize these features in the quantum regime by utilizing a sub-manifold of quantum states; transitions out of this sub-manifold are described by an effective non-Hermitian Hamiltonian. (c,d) The real and imaginary parts of the eigenvalue differences $\delta\lambda$ show an exceptional point at $J=\gamma_e/4$ along the $J$-axis (zero detuning). The yellow shaded region depicts the region explored in this experiment, with $\Delta = \omega_\mathrm{d} - \omega_\mathrm{q}$, where $\omega_\mathrm{d}$ is the drive frequency and $\omega_\mathrm{q}$ is the transition frequency between the $|e\rangle$ and $|f\rangle$ levels. (e) The experiment utilizes the three lowest levels of a transmon circuit. The circuit is embedded in a three-dimensional cavity and an Impedance Mismatch Element (IME) is used to shape the density of states that drive the decay of the transmon states through spontaneous emission.  A Josephson Parametric Amplifier (JPA) is used for high fidelity readout of the transmon state. Different decay rates can be obtained by threading a dc magnetic flux through the SQUID loop which tunes the frequency of the transmon energy levels.} 
\label{fig1}
\end{figure*}

A canonical example of a \pt-symmetric system consists of a bipartite system with balanced gain (part A) and loss (part B) as shown in Figure 1a. Such systems have been experimentally studied in the classical domain. The central feature of these systems is a transition from broken to unbroken \pt symmetry. When the coupling, given by rate $J$, between the two parts is larger than the gain-loss rate, given by \new{$\gamma_\pm$}, the system exhibits a real spectrum and simultaneous eigenmodes of both the Hamiltonian and the antilinear \pt operator; yet when the coupling is small, this \pt symmetry is broken by the emergence of complex conjugate eigenvalues. These two phases are joined by an exceptional point. The exceptional point degeneracy also occurs for a bipartite system with imbalanced losses. Figure \ref{fig1}b schematically displays such a system in which part A and B are coupled and part B exhibits loss. \new{Here we extend these studies to a fully quantum limit where these parts are realized as quantum energy levels---with no classical counterpart---where the loss corresponds to transitions outside that manifold of states.} This two-level system in the presence of coupling produced by a drive with strength $J$ and detuning $\Delta$ can be described by an effective non-Hermitian Hamiltonian ($\hbar = 1$), 
\begin{align}
    H_\mathrm{eff}=J\left(|f\rangle\langle e|+|e\rangle\langle f|\right)+(\Delta-i\gamma_e/2)|e\rangle\langle e|
\label{H_eff}
\end{align}
where $|e\rangle$ and $|f\rangle$ denote first and second excited states of the quantum system, $\gamma_e$ is the occupation-number loss rate to the ground state $|g\rangle$ (Fig.~\ref{fig1}b). At zero detuning, the complex eigenvalues of $H_\mathrm{eff}$ have different imaginary components at small $J<\gamma_e/4$ and the system is in the \pt-broken phase. At stronger coupling, past the exceptional point at $J=\gamma_e/4$, the imaginary components for the two dissipative eigenmodes coincide, and the system is in the \pt-symmetric phase. When $\Delta\neq 0$, the two complex eigenvalues $\lambda_{\pm}$ of the Hamiltonian $H_\mathrm{eff}$ (Eq.~1) have different real and imaginary parts. Here,  the qubit dynamics is governed by eigenmode-energy differences $\mathrm{Re}[\delta\lambda(\Delta,J)]$ (Fig.~\ref{fig1}c) and $\mathrm{Im}[\delta\lambda(\Delta,J)]$ (Fig.~\ref{fig1}d),  where $\delta\lambda=(\lambda_{+}-\lambda_{-})=\sqrt{4J^2-(\Delta-i\gamma_e/2)^2}$. 




\textit{Experimental setup}---Our experiment comprises a transmon circuit \cite{koch07} formed by a pair of Josephson junctions in a SQUID geometry shunted by a capacitor (Fig.~1e). The transmon circuit exhibits several quantum energy levels that can be individually addressed with narrow bandwidth microwave pulses. By applying a magnetic flux through the SQUID loop we can tune the spacing between energy levels. The coupling Hamiltonian $J\sigma_x$ is realized by a coherent resonant drive of variable amplitude and detuning.

The transmon circuit is embedded in a three-dimensional waveguide cavity \cite{paik113D}. The dispersive interaction between the transmon circuit and fundamental electromagnetic mode of the cavity results in a state dependent shift in the cavity frequency \cite{wall05}. This frequency shift is detected by probing the cavity with a weak microwave tone; the resulting state dependent phase shift is detected with homodyne measurement using a Josephson parametric amplifier \cite{cast08,hatr11para}.  The lowest energy level $|g\rangle$ is the stable ground state and we use it as an effective continuum---an environment that is ``outside"  of the sub-manifold of states $|e\rangle$ and $|f\rangle$ which form the qubit system under investigation. In order to implement the effective non-Hermitian Hamiltonian we require the respective energy decay rates $\gamma_e \gg \gamma_f$. \new{The presence of a finite decay rate $\gamma_f$ shifts the EP to $J =\gamma/4= (\gamma_e-\gamma_f)/4$. We achieve this hierarchy of decay rates by inserting an impedance mismatching element (IME)} between the cavity and parametric amplifier which causes an interference in the cavity field alternately suppressing and enhancing the density of states in the transmission line resulting a frequency dependence of the Purcell decay rate. Thus by tuning the transition frequency between the $|g\rangle$ and $|e\rangle$ states to regions where the density of states is enhanced, we enhance the decay rate of the $|e\rangle$ state.

\begin{figure*}
\centering
\includegraphics[width = .8\textwidth]{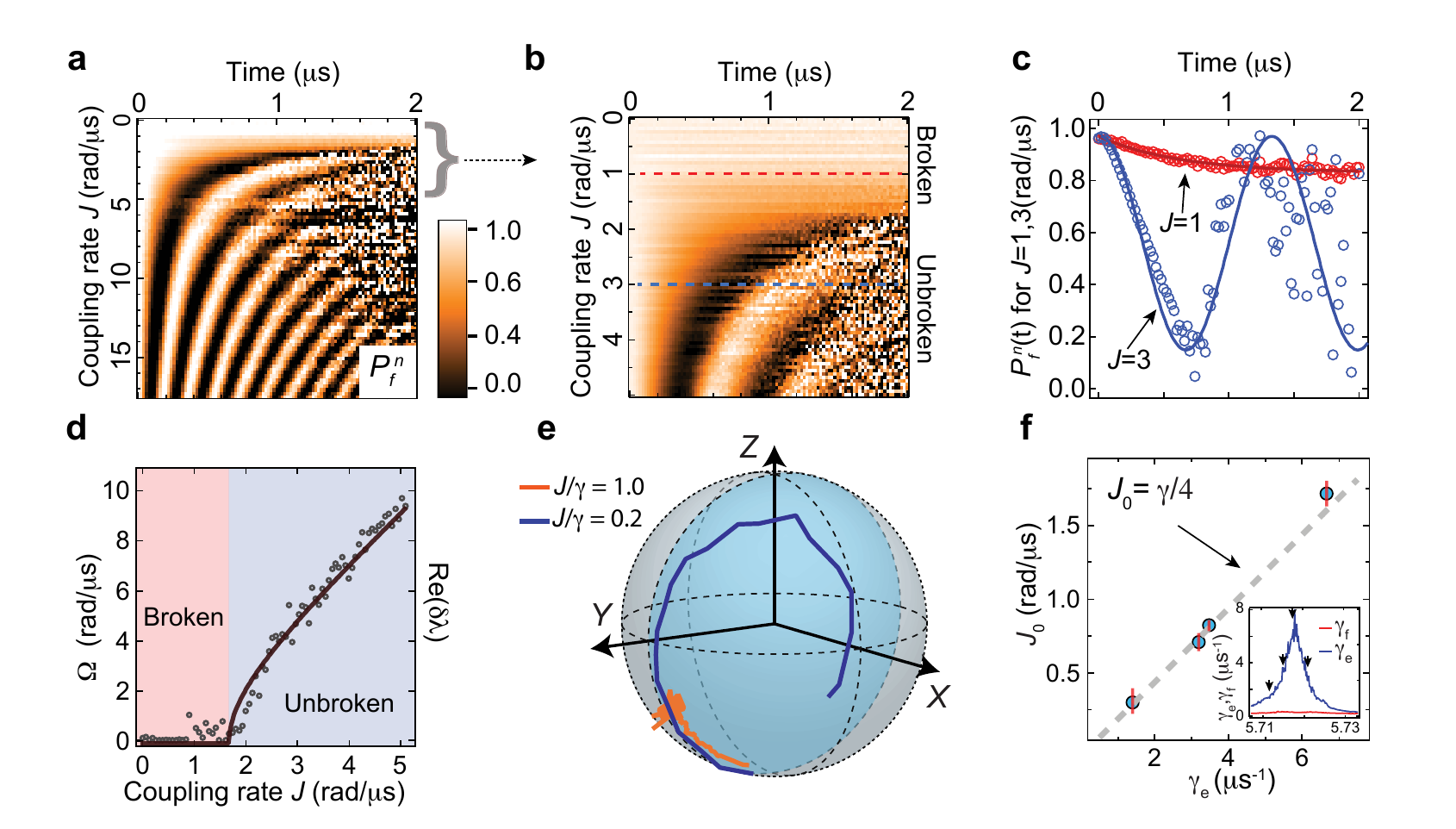}
\caption{{\bf \pt-symmetry breaking transition in a single dissipative qubit.} (a) Color map of the normalized population $P_f^n$ versus time for various coupling rates. \new{(b) Detail of $P_f^n$ for small values of $J$ highlighting the transition from the \pt-broken to \pt-symmetric phase. (c) Line cuts from panel (b) versus time show oscillatory to steady-state behavior.} (d) The extracted oscillation frequency $\Omega$ for different values of the coupling $J$.  The solid line indicates a fit to analytical result $\Omega=\mathrm{Re}\,\delta\lambda$. (e) Evolution of the quantum state in the Bloch sphere for parameters in the \pt-broken region ($J/\gamma = 0.2$, red) for time $t=[0,2]\ \mu$s, and in the \pt-symmetric region ($J/\gamma = 1.0$, blue) for time $t=[0,0.4]\ \mu$s. Both time intervals correspond to the same scaled time $2Jt$. (f) By tuning the transition frequencies of the transmon, different decay rates $\gamma_e$ can be obtained (inset).  The \pt-symmetry breaking threshold is obtained as in panel (d)  for four different values of the decay rate, showing good agreement with \mn{$J_0 = \gamma/4$ (dashed line). } }
\label{fig2}
\end{figure*}


\textit{\pt-transition and quantum state tomography}---We first investigate the \pt symmetry breaking transition which occurs when $\Delta = 0$.  We tune the transmon such that $\gamma_e=6.7\ \mu$s$^{-1}$ and $\gamma_f=0.25 \ \mu$s$^{-1}$. We then initialize the system in the state $|f\rangle$ with $J=0$ and at time $t=0$ we switch $J$ to a finite value for a variable period of time. The experimental sequence is concluded with a projective measurement of the transmon energy. 
\new{Evolution under $H_{\mathrm{eff}}$ leads to exponential decay of the norm of a given initial state. Experimentally, we focus on the evolution in the $\{|e\rangle,|f\rangle\}$ qubit manifold, which results in normalized populations, $P^{n}_f  =   P_f/(P_f+P_e)$ and $P^{n}_e  =   P_e/(P_f+P_e)=1-P^{n}_f$. }This is achieved through post-selection; experimental sequences conclude with a projective measurement of the transmon in the energy basis and only experiments where the transmon remains in the qubit manifold are included in the analysis. Thus, for longer experimental duration the success rate decreases exponentially. 

We now characterize the \pt-symmetry breaking transition using the observed experimental signatures in the populations and coherences in the $\{|e\rangle,|f\rangle \}$ qubit manifold. In Figure~\ref{fig2}a we show the normalized population $P^{n}_f$ versus time for different coupling rates $J$. For a large $J$ we observe oscillatory dynamics in $P^{n}_f$. These Rabi oscillations occur because the initial state $|f\rangle$ can be expressed as a superposition of eigenmodes of $H_\mathrm{eff}$ with corresponding time evolution $e^{-i\lambda_{\pm}t}$; the equal imaginary parts of $\lambda_\pm$ for $J>\gamma/4$ result in the oscillatory evolution at angular frequency $\Omega$ for the post-selected occupation probabilities.  This region is referred to as the \pt-symmetric region. The time evolution of $P^{n}_f$ shows a striking transition at finite coupling rate as detailed in Figure~\ref{fig2}b.  Here, we observe that when $J<\gamma/4$ the oscillations cease due to the vanishing real parts of $\lambda_{\pm}$.  This is referred to as the \pt-symmetry broken region. Figure \ref{fig2}c displays time-trace cuts from \ref{fig2}b in the broken and unbroken \new{regions} with decaying and oscillatory behavior respectively. \new{Although Figure 2 only displays experimental data where the transmon did not leave the $\{|e\rangle,|f\rangle \}$ qubit manifold, the post-selection on the qubit manifold leads to measurement backaction favoring the $\ket{f}$ state, leaving a clear signature of the decay in the temporal evolution within this manifold.}


The  \pt symmetry breaking transition can be quantified by looking at the oscillation frequency, $\Omega$, as a function of coupling rate. This oscillation frequency is obtained from a simple exponentially damped sinusoidal fit to $P_f^n(t)$ (Fig.~\ref{fig2}c). In Figure~\ref{fig2}d  we plot the observed oscillation frequency $\Omega$ versus coupling rate $J$, which displays a square-root singularity that is associated with increased sensitivity near the EP \cite{chen17,hoda17,chen18,lau18,meng18}.  The solid curve displays a fit to $\mathrm{Re}\,\delta\lambda= 2\mathrm{Re}\sqrt{J^2-J_0^2}$ with $J_0$ as the sole free parameter.  
\new{From the fit, we find $J_0=1.71\pm 0.07\ \mu$s$^{-1}$ which is in agreement with the expected value based on the independently measured decay rates $(\gamma_e-\gamma_f)/4=\gamma/4=1.61\ \mu$s$^{-1}$. }

Next we characterize the evolution of the qubit in the broken and unbroken regimes using quantum state tomography \cite{stef06}. Figure 2e displays $y\equiv \langle \sigma_y\rangle$ and $ z\equiv \langle \sigma_z\rangle$ (the initial state and Hamiltonian confine the evolution to the $Y$--$Z$ plane of the Bloch sphere) versus time for two different experimental conditions. While evolution in the \pt-symmetric phase shows oscillatory behavior, in the \pt-broken phase the state approaches a fixed point in the $Y$-$Z$ plane.  Both state trajectories are plotted for the same scaled time interval,  $0\leq 2J t\leq 5.24$ rad, highlighting the difference in quantum evolution in the symmetric and broken  phases.

We repeat the experiment for different values of \new{$\gamma$} by tuning the flux threading the transmon SQUID loop thereby and placing the transmon levels in contact with different parts of the engineered bath as depicted by arrows in Figure~\ref{fig2}f (inset). Figure~\ref{fig2}f shows the result from four different experiments. The \pt transition as determined from fits of the oscillation frequency for different $J$ as in Figure~\ref{fig2}d is in close agreement with the analytical \mn{result $J_0 = \gamma/4$.}



\begin{figure}
\centering
\includegraphics[width = .45\textwidth]{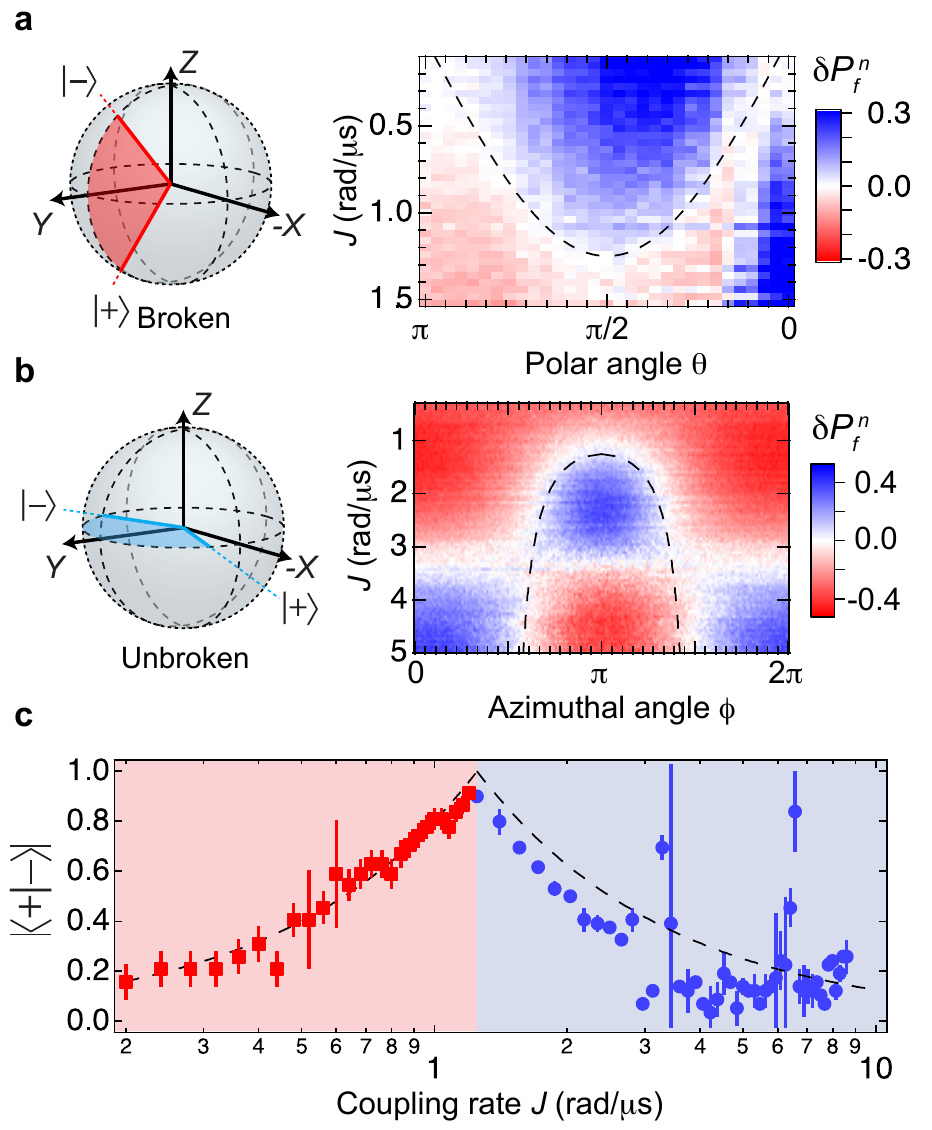}
\caption{\new{{\bf Non-orthogonality of eigenstates in the vicinity of the EP.}  The fractional change in ($P_f^n$) for different polar (a) and azimuthal (b) preparation angles. The calculated angles for the eigenstates in the broken (a) and unbroken (b) regions are indicated as dashed lines. (c)  The overlap between the two eigenstates in both regions satisfies $|\langle +|-\rangle |=\min(x,1/x)$ where $x=4J/\gamma$.}}
\label{fig3_orth}
\end{figure}

\new{Figure \ref{fig3_orth} we study the locations of the eigenstates of $H_\mathrm{eff}$ on the Bloch sphere as the system traverses the $\mathcal{PT}$ transition at the second-order EP.  We prepare different states of the qubit given by polar ($\theta$) and azimuthal ($\phi$) angles on the Bloch sphere. In the broken region (Fig.\ \ref{fig3_orth}a) the eigenstates appear as places where $\delta P_f^n = P_f^n(t=0)-P_f^n(t=500\ \mathrm{ns})$ is zero for different initial preparations in the $Y$--$Z$ plane. For the unbroken region (Fig.\ \ref{fig3_orth}b) these stationary states appear on the $X$--$Y$ plane. The expected stationary states, based on diagonalization of $H_\mathrm{eff}$ are given by dashed lines. The non-orthogonality of the eigenstates across the $\mathcal{PT}$ transition, including in the vicinity of the EP, is characterized in terms of the overlap $|\langle +|-\rangle |$ of the two eigenstates, displayed in Figure  \ref{fig3_orth}c, where the dashed line indicates the theoretical value, $|\langle +|-\rangle |=\min(x,1/x)$ where $x=4J/\gamma$.} 


\begin{figure*}
\centering
\includegraphics[width=0.9\textwidth]{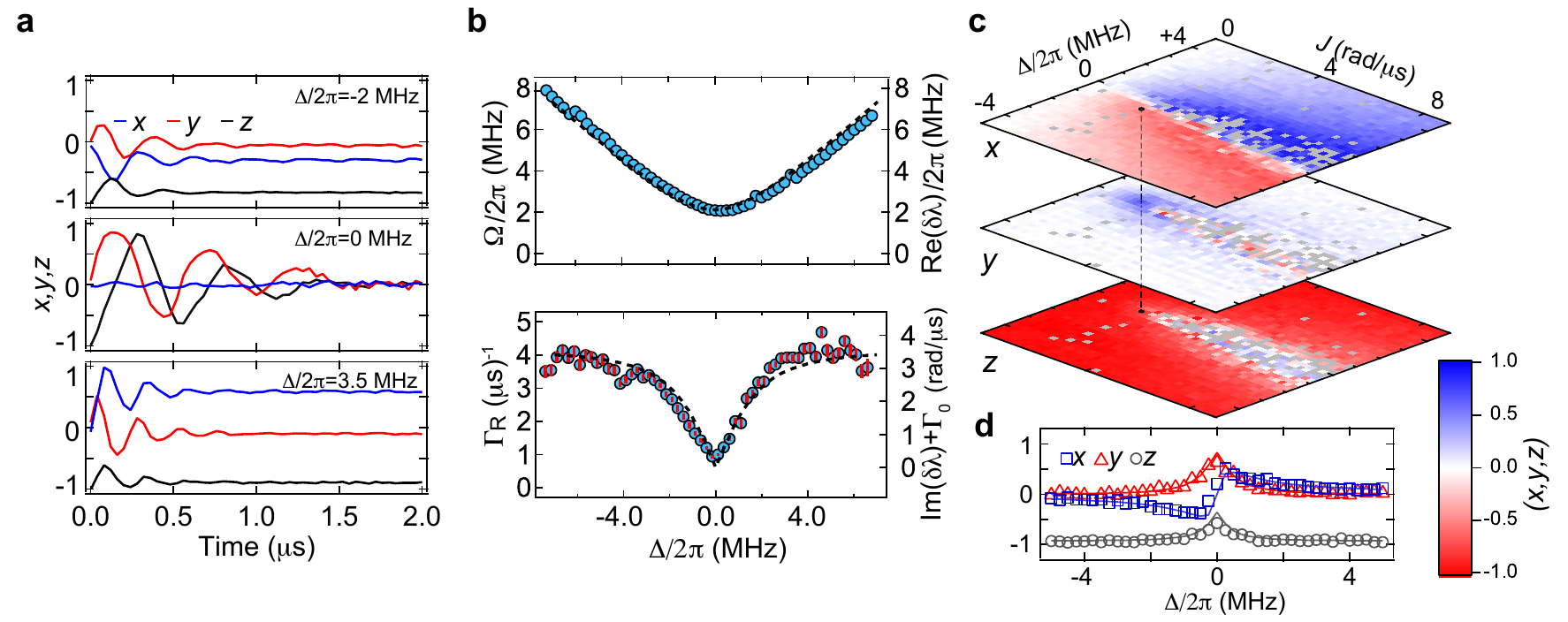}
\caption{{\bf Coherence damping and steady state of a single dissipative qubit.} (a) Time evolution of the Bloch components are fit to decaying sinusoidal curves to extract the oscillation frequency $\Omega$ and coherence-damping rate $\Gamma_R$ for different detuning values. (b) Observed Rabi frequencies and coherence-damping rates versus detuning in the \pt-symmetric region for $J = 6.9$ rad/$\mu$s and $\gamma_e = 7.1\ \mu$s$^{-1}$. The top-panel dashed line is \new{the} analytical result $\mathrm{Re}\,\delta\lambda$; the bottom-panel dashed line is the analytical result, \new{offset by the the residual coherence-damping term, $\mathrm{Im}\,\delta\lambda+\Gamma_0$}. (c) Quantum state tomography of the qubit for an evolution time \new{$t=4\ \mu$s} shows the steady states reached for different parameter regimes. Gray points indicate data points where there were an insufficient number of successful post-selections. \new{(d) A line-cut across the EP as a function of $\Delta$ shows a $y$ coherence bump that reaches unity when $\gamma_f=0$ (Methods)}.} 
\label{fig3}
\end{figure*}

{\it Decoherence in the vicinity of the EP}---With access to the quantum coherent dynamics in the vicinity of the exceptional point it is natural to investigate the role of decoherence in this regime. As shown in Figure~\ref{fig1}c,d the eigenvalue difference $\delta\lambda$ of $H_\mathrm{eff}$ exhibits rich dependence on $J$ and $\Delta$, which in turn determines the time evolution of the dissipative qubit.  Figure~\ref{fig3}a depicts the time evolution of the qubit state given by Bloch coordinates $x(t)$, $y(t)$, $z(t)$, which were measured with quantum state tomography for different values of the detuning. In the \pt-symmetric phase, we fit the oscillations to determine both the oscillation frequency $\Omega$ and the coherence damping rate $\Gamma_\mathrm{R}$ for different detunings, yielding respectively the real and imaginary parts of $\delta\lambda$ (Fig. \ref{fig3}b). At $\Delta=0$, the eigenmode decay rates are equal and we observe only a residual, small coherence-damping in the qubit manifold, characterized by $\Gamma_0= 0.6\  \mu\mathrm{s}^{-1}$; \new{this damping is larger than expected from the small $\gamma_f$ and is primarily due to charge and flux noise.} As $|\Delta|$ is increased, the difference in the eigenmode decay rates leads to faster coherence damping.  \new{The observed $\Omega$ and $\Gamma_\mathrm{R}$ are in good agreement with the analytical predictions offset by the residual zero-detuning coherence damping $\Gamma_0$}. 

Quantum state tomography also allows us to study the steady states of the qubit system evolving under $H_\mathrm{eff}$ in the vicinity of the exceptional point. Figure~\ref{fig3}c displays the steady-state results of quantum state tomography after \new{4 $\mu$s} of time evolution. Along the \pt-symmetric phase line ($\Delta=0$ and \mn{$J>\gamma/4$}), the qubit reaches a maximally mixed state. When $|\Delta|>0$, the qubit reaches a mixed steady state in the $X$--$Z$ plane, i.e. $y\sim 0$. Remarkably, in the close proximity of the EP, \new{when $\gamma_f\rightarrow 0$}, the qubit reaches a steady state given by $(|e\rangle+i|f\rangle)/\sqrt{2}$, i.e. the single eigenmode of $H_\mathrm{eff}$ at the EP. \new{In our experiment, this appears as a peak in the $y$-component in the tomography in Figure~\ref{fig3}c, along with vanishing $x$ component, and a $z$ component that is suppressed in magnitude.} These results indicate that the dissipation of the system stabilizes the qubit to non-trivial steady-states for different drive and detuning parameters.



\new{\textit{Outlook}---While the dynamics of the 3-level transmon are described by a Lindblad equation with two dissipators that characterize spontaneous emission from levels $\ket{f}$ and $\ket{e}$, the non-Hermitian evolution and EP effects are only manifest when quantum jumps to the $\ket{e}$ state are eliminated by post selection~\cite{MolmerPRL92}}. 
\new{Using this approach we have explored the EP signatures in the quantum domain by investigating the quantum coherent dynamics in its vicinity. } These results highlight how circuit quantum electrodynamics serves as a versatile platform to explore fundamental questions in the quantum mechanics of open systems. \new{Recent work identifying enhanced sensitivities in the vicinity of the EP have spurred interest in the role of quantum noise in EP-based sensors~\cite{lau18,meng18,chen18}.  Our system forms an ideal platform characterizing quantum sensing applications using non-Hermitian systems including the role of noise entering from dissipation (Methods). Finally, real time control over the parameters of the effective non-Hermitian Hamiltonian will allow studies of topological features associated with adiabatic perturbations that encircle the exceptional point, and of higher-order exceptional surfaces that arise in time-periodic (Floquet) non-Hermitian dynamics.}


\section{Methods}

In this section we provide details of the experimental setup and techniques utilized in this work. We also provide an analysis of the system as described by a Lindblad evolution in the three-state manifold, which is equivalent to the non-Hermitian Hamiltonian evolution in the two-state manifold. \new{We provide further discussion regarding the interplay of Lindbladian dissipation and non-Hermitian dissipation as well as prospects for enhanced sensing near the EP.} \\

\begin{figure}[t]
\centering
\includegraphics[width = 0.48\textwidth]{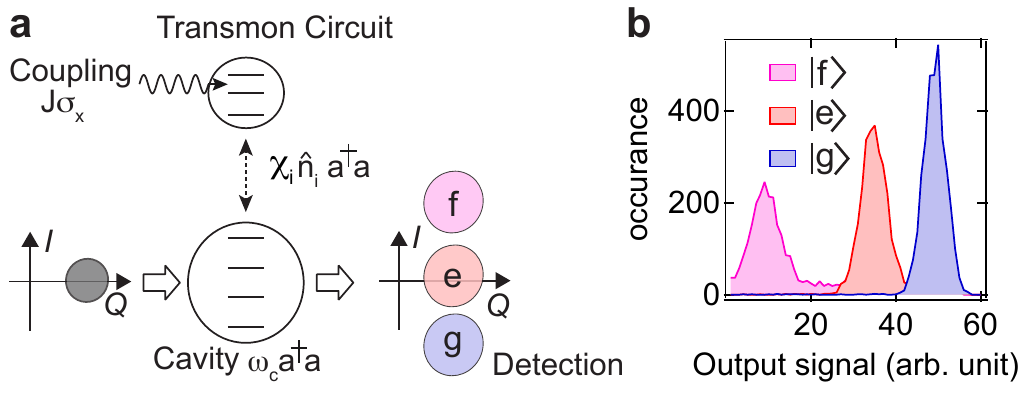}
\caption{State readout scheme. (a) Readout of the transmon circuit states is achieved through the dispersive interaction between the circuit and a cavity mode.  An input drive on the cavity acquires a transmon-state dependent phase shift which is amplified with a near-quantum-limited Josephson parametric amplifier. (b) After demodulation, the $I$ quadrature of the signal is integrated for 160 ns, resulting in well-separated measurement histograms.} 
\label{fig1_sup}
\end{figure}

\emph{Experimental setup}---The transmon circuit was fabricated by conventional double-angle evaporation and oxidation of aluminum on a silicon substrate. With zero flux threading the SQUID loop, the  transition frequencies are $\omega_{g,e}/2\pi= 6.1$ GHz and $\omega_{e,f}/2\pi= 5.8$ GHz. The transmon circuit is placed in a 3D copper cavity with frequency $\omega_\mathrm{c,bare}/2\pi = 6.681$ GHz and decay rate $\kappa/2\pi =5$ MHz with an embedded coil for adjusting the dc magnetic flux through the SQUID loop. The coupling rate between the transmon circuit and the cavity fields is $g/2\pi=65$ MHz. Experiments are performed with a small flux threading the SQUID loop resulting in transition frequencies, $\omega_{g,e}/2\pi \simeq 5.71$ GHz and $\omega_{e,f}/2\pi \simeq 5.42$ GHz, given by charging energy $E_\mathrm{c}/h = 270$ MHz and Josephson energy $E_\mathrm{J}/h = 16.6$ GHz where the dressed cavity resonance frequency is $\omega_\mathrm{c}/2\pi = 6.684$ GHz and the dispersive cavity resonance shifts are given by $\chi_{e}/2\pi = -2$ MHz and $\chi_f/2\pi = -11$ MHz. In order to rapidly resolve the transmon states with high fidelity, we use a Josephson parametric amplifier operating in phase sensitive mode with 20 dB gain and instantaneous bandwidth of 50 MHz. As shown in Figure \ref{fig1_sup} we are able to resolve the three transmon states with high fidelity.

\new{\emph{Data analysis and experimental error}---In Figure \ref{fig3_orth} we extract the locations of the eigenstates in the broken and unbroken regions.  This is achieved through a two point measurement technique. In the unbroken region, the eigenstates are simply found by comparing the change in $P_f^n$ over 500 ns of evolution.  States that are stationary exhibit no change, whereas non-eigenstates exhibit oscillatory behavior.  In the broken regime, although the eigenvalues are strictly imaginary, the stationary states are still visible as regions where $P_f^n$ is stationary. The data displayed in Figure 3a have been scaled to account for the small $\gamma_f$ decay over 500 ns. The preparation angles for the eigenstates were found from the zero crossing of the $\delta P_f^n$ plots, determined from $\mathrm{min}[\mathrm{abs}(\delta P_f^n)]$, and the error bars indicate the distance to the next-nearest minima.  For this data set $\gamma_e = 5.25 \ \mu\mathrm{s}^{-1}$ and $\gamma_f = 0.25 \ \mu\mathrm{s}^{-1}$.}

\emph{Lindblad evolution of the three-state system}---In the main text, we solely focused on the dynamics in qubit subsystem which is governed by the effective, dissipative Hamiltonian $H_{\mathrm{eff}}$, Eq.~(\ref{H_eff}). Instead, one can look at the dynamics for the entire 3-level system which can be described by a Lindblad master equation ($\hbar=1$),
\begin{eqnarray}
\dot{\rho} = -i [ H_c, \rho ] +  \sum_{k=e,f}\left[L_k\rho L_k^{\dagger} - \frac{1}{2} \{L_k^{\dagger}L_k , \rho \}\right].
\label{linblad}
\end{eqnarray}
\mn{Where $\rho(t)$ is a $3\times3$ density matrix, $H_c = J ( |e\rangle \langle f| + |f\rangle \langle e|) + \Delta (|f\rangle \langle f| - |e\rangle \langle e|)$ is coupling Hamiltonian with detuning $\Delta$ in the rotating frame. The Lindblad dissipation operators $L_e=\sqrt{\gamma_e} |g\rangle \langle e|$ and $L_f=\sqrt{\gamma_f} |e\rangle \langle f|$ account for the energy decay from level $|e\rangle$ to $|g\rangle$ and $|f\rangle$ to $|e\rangle$ respectively. Equation~(\ref{linblad}) leads to the following closed set of equations for the dynamics of the qubit levels,
\begin{eqnarray}
\dot{\rho}_{ff} &=& -i J (\rho_{ef} - \rho_{fe})  - \gamma_f \rho_{ff} \nonumber \\
\dot{\rho}_{ee} &=& +i J (\rho_{ef} - \rho_{fe})  - \gamma_e \rho_{ee} + \gamma_f \rho_{ff} \nonumber\\
\dot{\rho}_{ef} &=& -i J (\rho_{ff}-\rho_{ee})  - (\gamma_e+\gamma_f+2\Delta)/2 \rho_{ef} \nonumber\\
\dot{\rho}_{fe} &=& +i J (\rho_{ff}-\rho_{ee})   - (\gamma_e+\gamma_f-2\Delta)/2 \rho_{fe}.
\label{ODEs}
\end{eqnarray}
}
Since the drive only acts on the manifold of two excited states, the dynamics of the ground state is decoupled from the upper manifold. For a given initial condition, one can solve Eqs.~(\ref{ODEs}) and obtain the evolution of any observable. As in the experiment, where the system is initialized in the state $|f\rangle$, \mn{and in the limit of $\gamma_f\ll \gamma_e$ and $\Delta=0$,} the evolution for the populations of each level in the \pt-symmetric phase is given by,
\begin{eqnarray}
P_e =\rho_{ee}&=& e^{-\frac{\gamma_e}{2} t} (\frac{J}{\alpha})^2   \sin^2( \alpha t ) \label{pe}\\
P_f =\rho_{ff}&=& e^{-\frac{\gamma_e}{2} t} (\frac{J}{\alpha})^2 \cos^2(\alpha t- \theta) \label{pf}
\end{eqnarray}
where $\alpha=\sqrt{J^2- (\gamma_e/4)^2}$ and $\theta=\arcsin(\gamma_e/4J)$.

In the main text, all analysis is performed in a model-independent manner; the evolution of the post-selected occupation number $P_f^n(t)$ is fit to an exponentially decaying sine function to determine the coherence-decay rate and the Rabi oscillation frequency. With access to the exact evolution in the three state system we can determine the actual form for the oscillation in the sub-manifold (e.g. Fig~\ref{fig2}c). From Eq.~(\ref{pe}),(\ref{pf}) we can obtain the normalized population,
\begin{eqnarray}
P^n_f = \frac{P_f}{P_f+P_e} =  \frac{ \cos^2(\alpha t -\theta)}{ \sin^2(\alpha t) + \cos^2(\alpha t -\theta)}. \label{eq:exact}
\end{eqnarray}
In the limit of $J \gg \gamma_e\gg\gamma_f$, Eq.~\eqref{eq:exact} reduces to $\cos^2(Jt)$, which means \new{that} deep in the \pt-symmetric region, far away from the EP, the population oscillates with frequency of $2J$. The observed oscillation frequency at $J\gg \gamma_e$ was used to calibrate the values of $J$ for weaker drives. These results are consistent with the direct theoretical approach for the evolution of the qubit wave function under non-Hermitian Hamiltonian $H_\mathrm{eff}$.

\new{\emph{Quantum state tomography in the vicinity of the EP}---Figure 4c displays quantum state tomography for a fixed evolution time $t=4\ \mu$s as a function of $\Delta$ and $J$.  At $t=4\ \mu$s the number of successful post-selections can be quite low, especially at $\Delta=0$, where the evolution takes the qubit through the lossy $\ket{e}$ state. }

\begin{figure}[h]
\centering
\includegraphics[width = 0.48\textwidth]{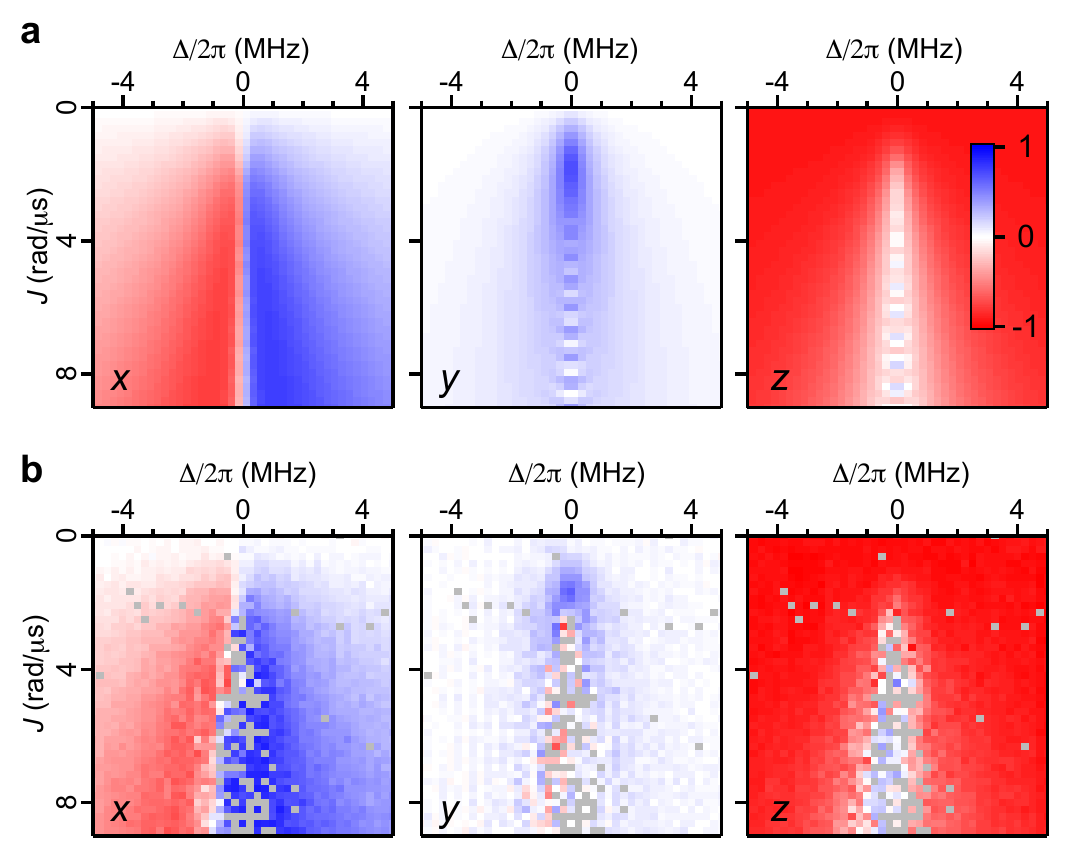}
\caption{\mn{The steady state tomography for different detuning $\Delta$ and coupling rate $J$. (a) Simulation using numerical solution to Equations \ref{ODEs}. (b)  Experimental result from Figure 4c.}}
\label{fig:tomo}
\end{figure}

\new{Figure \ref{fig:tomo} displays comparisons of the tomography data to simulations using Eqs. \ref{ODEs}, for the same evolution time $t=4\ \mu$s. We note oscillations for $\Delta=0$ have not completely damped out for this evolution time. We attribute the faster damping in the experimental data to additional dephasing, characterized by $\Gamma_0$ which was not included in the simulation.  Otherwise we see good qualitative agreement between simulation and experimental data.}

\new{We also measured $\Gamma_0$ for a different flux bias of the transmon where $\gamma_e = \gamma_f = 0.14\ \mu\mathrm{s}^{-1}$, and found $\Gamma_0=0.46 \ \mu\mathrm{s}^{-1}$ for $J = 6.9 \ \mathrm{rad}/\mu\mathrm{s}$ in fairly close agreement to what was observed in Figure 4. From this we conclude that the additional dephasing is likely due to flux or charge noise in the transmon and not an additional feature of the effective non-Hermitian evolution. }

\begin{figure}[h]
\centering
\includegraphics[width = 0.48\textwidth]{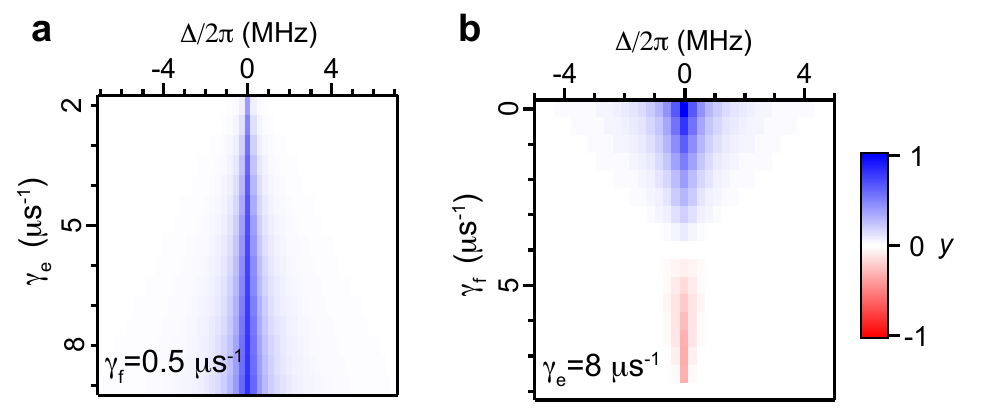}
\caption{\new{(a) The steady state $\langle \sigma_y\rangle$ at values of the coupling rate corresponding to the EP for $\Delta=0$ versus $\Delta$ determined by solving the Lindblad master equation (\ref{ODEs}) for different values of $\gamma_e$ and fixed $\gamma_f$.  (b) Steady state $\langle \sigma_y\rangle$ versus $\gamma_f$ for fixed $\gamma_e$.}}
\label{fig:bump}
\end{figure}

\new{\emph{Interplay between non-Hermitian ($L_e$) and Lindbladian ($L_f$) dissipation}---A remarkable feature that quantum state tomography uncovered in Figure \ref{fig3}c is that the combination of non-Hermitian evolution and dissipation produces a steady state of the qubit along the $+\hat{y}$ axis.  Here we examine this feature through simulations of the Lindblad master equation for the three state system where both $L_e$ and $L_f$ are present with comparable magnitudes. In Figure \ref{fig:bump}a we display the steady state $y$ as a function of $\Delta$ for $J = (\gamma_e-\gamma_f)/4$, corresponding to the EP for $\Delta = 0$.  We observe that while finite $\gamma_f$ is necessary for the formation of a steady state, the steady state coherence is maximum for extremal ratios of $\gamma_e/\gamma_f$.  Figure \ref{fig:bump}b displays a similar calculation, but for different values of $\gamma_f$.  We observe that at $\gamma_f = \gamma_e/2$ the steady state $y$ changes sign, approaching that expected for a normal dissipative qubit where the balance of drive and decay can result in a steady state coherence \cite{carm87} with a negative $y$. This transition occurs when the Lindbladian dissipation overtakes the non-Hermitian dissipation, which occurs at $\gamma_f = \gamma_e/2$ }

\new{\emph{Quantum sensing in the vicinity of the EP}---Recent work with classical systems has indicated EP degeneracies may yield measurement advantages~\cite{wier14,hoda17,chen18}. These studies have motivated further investigation into whether these advantages persist in the fully quantum regime where quantum noise dominates the measurement process. Theoretical work on semi-classical optical systems~\cite{lau18,meng18} has found that enhanced sensitivities near the EP are counteracted by enhanced fluctuations, curtailing measurement advantages. How these studies extend to the fully quantum regime explored here remains an open question.  In this section, we briefly discuss how the Lindblad evolution of the 3-state system can be used to characterize enhanced measurement sensitivities in terms of the quantum Fisher information (QFI), as well as how the post-selection process may hamper these advantages.}

\new{In quantum metrology, the Cram\'er-Rao bound \cite{cram46} gives a universal limit for the mean squared deviation an estimate of a parameter, 
\begin{align}
\langle \delta^2 {\hat g} \rangle \geq \frac{1}{v I_g^{(Q)}}, 
\end{align}
where $v$ is a measure of the amount of data, ${\hat g}$ is an unbiased estimator of the parameter $g$ formed from measurement data, and $I_g^{(Q)}$ is the quantum Fisher information, which can be expressed in terms of the Bures distance \cite{bure69}, $\mathrm{d}s^2 = 2(1-|\langle \psi_{g}|\psi_{g + \mathrm{d}g}\rangle |)$, where $I_g^{(Q)} = 4  \mathrm{d}s^2/\mathrm{ d}g^2$.}

\begin{figure}[h]
\centering
\includegraphics[width = 0.48\textwidth]{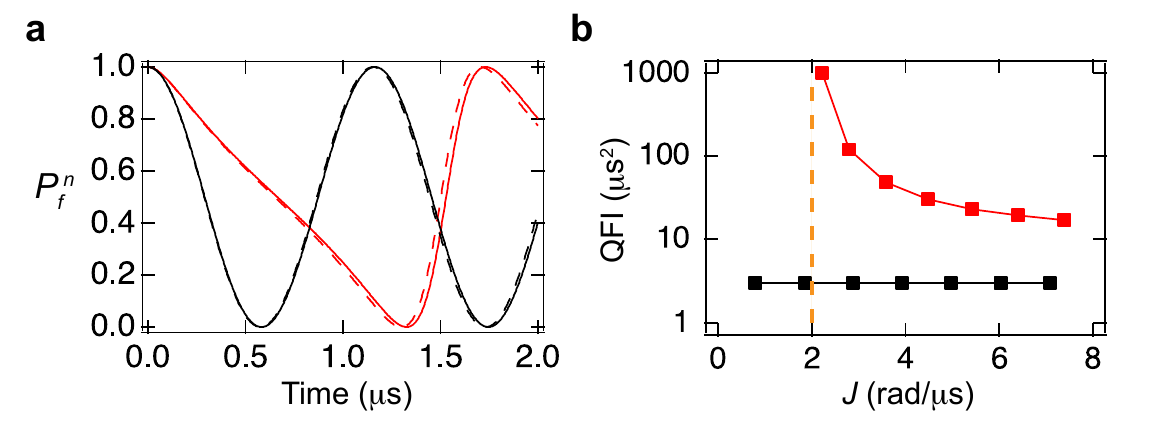}
\caption{\new{(a) Calculation of the $P_f^n$ versus time for $J = 2.7 \ \mathrm{rad}/\mu\mathrm{s}$ and $\gamma_e = 8\ \mu\mathrm{s}^{-1}$ (red) and $\gamma_e = 0\ \mu\mathrm{s}^{-1}$ (black), with $\gamma_f = 0$ in both cases.   The dashed lines indicate $0.7$ \% changes in $J$.  (b) The QFI in the qubit manifold versus $J$ for $\gamma_e = 8\ \mu\mathrm{s}^{-1}$ (red) and $\gamma_e = 0\ \mu\mathrm{s}^{-1}$ (black). The location of the EP is indicated as a dashed orange line.  The QFI in the qubit manifold about the coupling $J$ diverges near the EP.}}
\label{fig:qfi}
\end{figure}

\new{One approach to metrology near the EP is based on Rabi interferometry. For this, we consider preparing the qubit in state $\ket{f}$, and allowing evolution under $H_\mathrm{eff}$ for certain durations of time.  Figure \ref{fig:qfi}a displays the evolution of $P_f^n$ for parameter regimes that are near the EP, calculated using the Equations \ref{ODEs} for the 3 state system.  The evolution near the EP is not purely sinusoidal and we note that there are points where the $f$-state population varies rapidly with time. By changing $J$ by a small amount, we observe a large change in the $f$-state population compared to the case of a normal Hermitian qubit with no EP for the same evolution time. The fractional change in the $f$-state population for a fractional change in $J$ is closely related to the quantum Fisher information.}

\new{To determine the QFI, we simply vary $J$ by a small amount to determine the slope $\mathrm{d}P_f^n/\mathrm{d}J$.  For small changes near $P_f^n = 0.5$, we have $P_f^n-\frac{1}{2} = \cos(\mathrm{d}\theta/2)\sin(\mathrm{d}\theta/2) \simeq \mathrm{d}\theta/2$, where $\mathrm{d}\theta$ is a small change in polar angle near the equator of the Bloch sphere. Thus, near the equator of the Bloch sphere, the QFI about the coupling rate $J$ is simply given by  $I_J^{(Q)} = (\mathrm{d}P_f^n/\mathrm{d}J)^2$.}

\new{Figure \ref{fig:qfi}b displays the QFI for this measurement scheme near the EP based on the parameters used in Figure \ref{fig:qfi}a.  We note that the QFI diverges near the EP, as has been observed for the classical Fisher information in classical systems.  This improved QFI, however, comes at a cost due to the post-selection that is used to realize the effective non-Hermitian dynamics; near the EP, the post selection efficiency is low, which ultimately decreases the amount of data available.  In this way, the enhanced sensitivity near the EP bears similarities to weak value amplification, where low post-selection efficiency is at odds with amplified signals. We note that even in this case, advantages to post-selection remain when signals are dominated by technical noise \cite{jord14}. }


\new{\emph{Note}---During preparation of this manuscript we became aware of other recent work using superconducting circuits \cite{part18} and and an experiment with nitrogen vacancy centers \cite{wu18}.}

\emph{Acknowledgments}---We acknowledge P.M. Harrington for preliminary contributions, D. Tan for sample fabrication and K. M\o lmer and C. Bender for discussions. KWM acknowledges research support from the NSF (Grant PHY-1607156, and PHY-1752844 (CAREER)), and YJ acknowledges NSF grant DMR-1054020 (CAREER). This research used facilities at the Institute of Materials Science and Engineering at Washington University.

\emph{Author Information}  
Correspondence and requests for materials should be addressed to KWM and YJ (murch@physics.wustl.edu, yojoglek@iupui.edu).

\end{document}